\documentstyle[11pt,moriond,epsfig]{article}

\def\Journal#1#2#3#4{{#1} {\bf #2}, #3 (#4)}

\def\PLB{{\em Phys. Lett.}  B}

\def\PRD{{\em Phys. Rev.} D}

\def\be{\begin{equation}}
\def\ee{\end{equation}}
\def\bea{\begin{eqnarray}}
\def\eea{\end{eqnarray}}

\begin{document}
\vspace*{4cm}
\title{THE SCALAR MESON PUZZLE BEYOND BCS}

\author{ PEDRO BICUDO }

\address{Departamento de F\'\i sica, Instituto Superior T\'ecnico,
Av. Rovisco Pais,\\
1049-001 Lisboa, Portugal
}

\maketitle\abstracts{
We present a new perspective \cite{bic} on the scalar meson puzzle from
spontaneous breaking of chiral symmetry beyond BCS.
We find that going beyond BCS does not produce a systematic 
shift of the hadron spectrum.
We also show that coupled channels reduce the breaking of chiral
symmetry, with the same Feynman diagrams that appear in the
coupling of a scalar meson to a pair of pseudoscalar mesons.
With a Lattice QCD inspired quark-quark interaction, we find 
that the groundstate $I=0, \ {}^3P_0$ $q \bar q$ meson is 
the $f_0(980)$ with a partial decay width of $40MeV$.
We also find a $30 \%$ reduction of the chiral condensate
due to coupled channels.}
\begin{figure}
\begin{picture}(0,0)(0,0)
\put(165,340){\epsffile{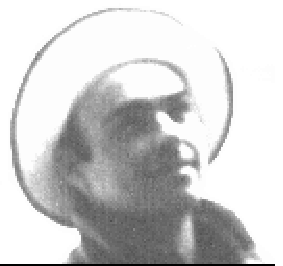}}
\put(165,340){\framebox(132.5,133){}}
\end{picture}
\end{figure}
\section{A link from QCD to Hadronic Physics}
We first integrate formally the gluons from the
QCD action, and get an action of Dirac quarks which 
interact via cumulants of gluons \cite{nora1}. This is the 
cumulant expansion,
\begin{eqnarray}
{\cal H}_{int}= { 1\over 2}\int \,d^4x\,d^4y\,
\overline{\psi}( x)
\gamma^\mu{\lambda^a \over 2}\psi ( x)  \; 
g^2 \langle A_\mu^a(x) A_\nu^b(y) \rangle
\;\overline{\psi}( y)
\gamma^\nu{\lambda^b \over 2}
 \psi( y)  \ + \ \cdots
\label{eq:Hint}
\end{eqnarray}
The first cumulant, of two gluons, can be evaluated 
in the modified coordinate gauge,
\begin{equation}
\langle A_\mu^a(x) A_\nu^b(y) \rangle
=x^ky^l\int^1_0 d \alpha d \beta
\alpha^{n(\mu)}\beta^{n(\nu)}
\langle F^a_{k \mu}(x_0,\alpha {\bf x})
F^b_{l \nu}(y_0,\beta {\bf y}) \rangle
\label{eq:cond}
\end{equation}
as a function of the non local gluon condensate which
can be evaluated on the lattice \cite{'t Hooft,Giacomo}. 
Other links from the Lattice to hadronic physics 
\cite{baker,brower} were presented at this conference.
If we use the approximation of including only the first
gluon cumulant in the Bethe-Salpeter kernel, we get
a relativistic quark model, where the bare hadrons 
are described by a ladder of a quarks (antiquarks).
Moreover, from the cumulant expansion we obtain
a chiral invariant interaction.
\par
Hadronic physics has 3 crucial properties
that we now must recover from this quark model.
Confinement is included from the onset in the
quark-quark interaction.
Spontaneous chiral symmetry breaking is achieved
when the mass gap equation for the quark is solved.
Strong interactions are finally recovered when
the couplings of bare hadrons are included.
\par
\section{The BCS level}
The BCS level corresponds to the minimal use of the
interaction both in the mass gap equation and in
the Bethe Salpeter equation \cite{bic}.
\subsection{A microscopic quark-quark interaction.
}\label{subsec:inter}
The calculation on the lattice \cite{Giacomo} of the full 
non local gluon correlator lends support \cite{nora1}
for the picture of a simplified instantaneous model,
\begin{equation}
\label{interaction}
g^2 \langle A_\mu^a(x) A_\nu^b(y) \rangle
\simeq{-3 \over 4} \delta_{ab} g_{\mu \nu} \left\{
g_{\mu 0}
\left[K_0^3({\bf x}-{\bf y})^2-U\right]
+ a g_{\mu i}k_0^3({\bf x}-{\bf y})^2
\right\}\delta\left(x^0-y^0\right) \ \ .
\end{equation}
In this interaction, a single scale $K_0$ is independent and
needs to be fixed with the spectroscopy. 
The best fit is obtained \cite{pene} with $K_0=330MeV$. 
The remaining parameters are determined from confinement 
($U\rightarrow \infty$) and Lorentz invariance of the 
$\pi$ ($a=-.18$). 
This single scale of hadronic physics can be measured 
for instance in the wave-functions of vector mesons
\cite{clerbaux}. 
\subsection{The mass gap equation and constituent quarks
}\label{subsec:mass}
\begin{figure}
\begin{picture}(403,150)(0,0)
\put(60,-26){\epsffile{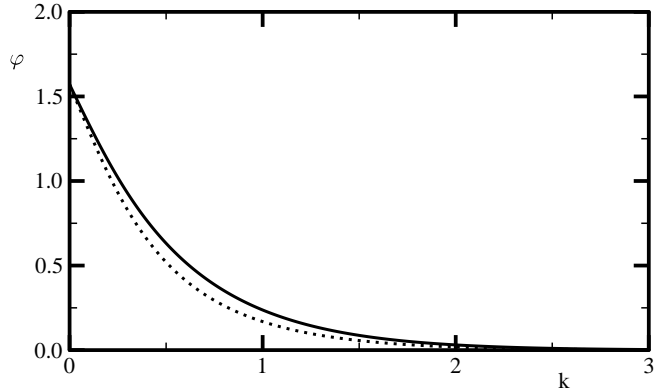}}\end{picture}
\caption{ We show with a solid line the BCS chiral 
angle $\varphi=\arctan\left[m(k) /k \right]$
in dimensionless units of $K_0=1.$
We also represent with a doted line the chiral angle
that we obtain going beyond BCS.
\label{chiral}}
\end{figure}

The mass gap equation is the Schwinger-Dyson equation,
where $V$ is the Fourier transform of the first cumulant,
\begin{equation}
{\cal S}^{-1}(p)={\cal S}_0^{-1}(p) - \Sigma \ , \ \
\Sigma =\int {d^4q \over (2\pi)^4}
\gamma^\mu{\lambda^a \over 2} {\cal S}(q)
V_{\mu \nu}(p-q)
\gamma^\nu{\lambda^b \over 2} \ \ .
\end{equation} 
For the interaction (\ref{interaction}) several solutions
for the Dirac quark propagator exist
\cite{pene,tese}. The solution for the
dynamical (constituent) quark mass $m$ which minimizes 
the vacuum energy density is illustrated in Fig. \ref{chiral}.
We also find that the energy density of the vacuum is 
infinite and this might be relevant for gravitation \cite{mathur}.
\subsection{The Bethe Salpeter equation and bare mesons
}\label{subsec:bare}
\par
The boundstates appear as poles in the ladder,
\begin{equation}
\label{ladder resum}
\begin{picture}(170,27)(0,0)
\put(0,0){\epsffile{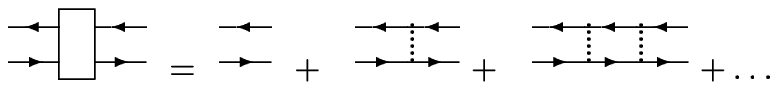}}\end{picture}
\end{equation}
The full hadron spectrum can be computed \cite{pene}. In particular 
we find $ M_{f_0}\simeq 1 \, GeV$ and a massless $\pi$ in the chiral
limit. 
\subsection{Coupled Channels}\label{subsec:chan}
The coupled channel contribution to the spectrum
is evaluated with the mesonic form factors. At the
quark level the form factors are computed with the overlaps,
\begin{equation}
\label{F(P)}
\begin{picture}(400,40)(0,0)
\put(0,0){\epsffile{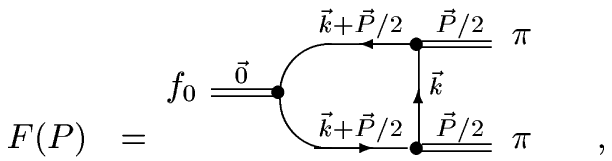}}\end{picture}
\end{equation}
where the blobs represent the vertex Bethe
Salpeter amplitudes.
\subsection{The scalar decay width}\label{subsec:wid}
We finally find for the scalar meson,
\begin{equation}
\Gamma_{f_0 \rightarrow \pi \pi}= 2i Im[M_{f_0}]=
{1 \over 4 \pi} Q M_{f_0}\left|F(Q)\right|^2\simeq \, 40 MeV 
\ , \ \
Q={1 \over 2} \sqrt{M_{f_0}^2-4M^2_\pi} \ \ .
\end{equation}
\section{Going beyond BCS}
However, when the coupled channels are included in
the boundstate equations, we should go beyond the BCS 
mass gap equation.
Otherwise the $\pi$ would get a negative contribution 
from the coupled channels, and it would have a paradoxal 
negative mass \cite{bic}.
\subsection{New mass gap equation}\label{subsec:nema}
Naively, when the coupled channels in the Bethe 
Salpeter equation for the vertex, only the first line 
of eq.(\ref{CS3 kernel}) is considered.
Using the vector Ward Identity,
\begin{equation}
\label{vector WI}
i (p_{\mu}-p_{\mu}') {\cal S}(p)\Gamma^{\mu}(p,p'){\cal S}(p')
= {\cal S}(p) -{\cal S}(p')   
\end{equation}
we replace the vertex by propagators and 
get the new quark self energy $\Sigma$,
\begin{eqnarray}
\label{CS3 self}
\begin{picture}(110,40)(0,0)
\put(0,0){\epsffile{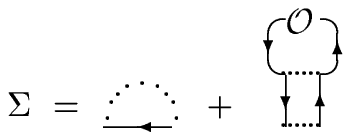}}\end{picture} \ , \ \
\begin{picture}(150,50)(0,0)
\put(0,0){\epsffile{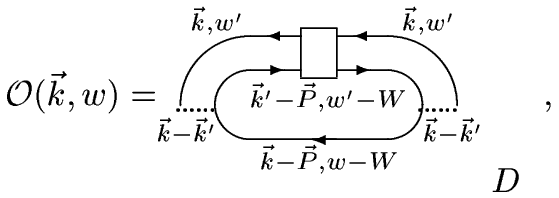}}\end{picture}
\end{eqnarray}
It turns out that a pseudoscalar loop and the coupling 
$F(p)$ of a scalar to a pair of pseudoscalars dominate
the new coupled channel term. This opposes to the breaking
of chiral symmetry. We show in Fig.[1] that the chiral angle
which measures the extent of spontaneous breaking of
chiral symmetry decreases.
\subsection{New Bethe Salpeter equation
}\label{subsec:nebe}
Inversely the new Bethe Salpeter equation, consistent with the new 
mass gap equation, is derived from eq.(\ref{CS3 self}) with the 
vector Ward Identity (\ref{vector WI}). 
Replacing a propagator by a vertex in all possible combinations we get,
\begin{eqnarray}
\label{CS3 kernel}
\begin{picture}(400,200)(20,0) 
\put(0,0){\epsffile{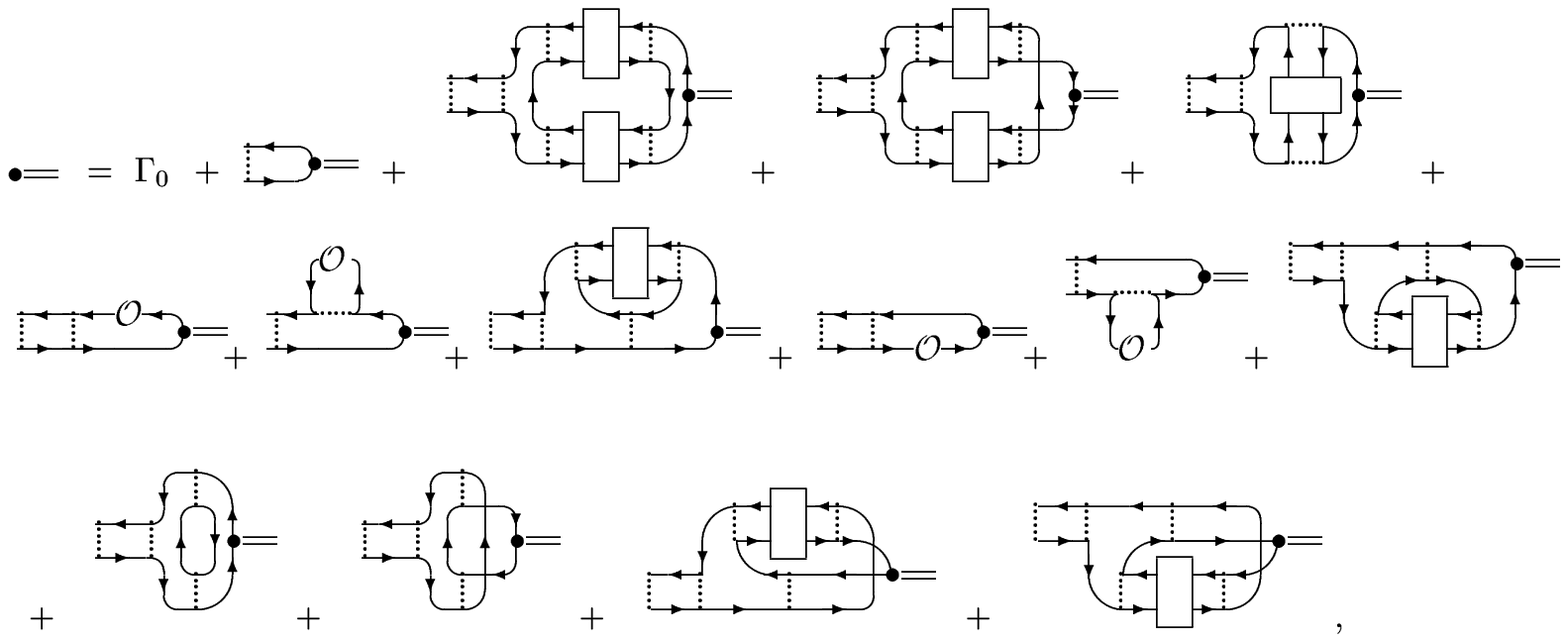}}\end{picture}
\end{eqnarray}
where in the first line we include the BCS terms
and the terms which are included in naive
coupled channel studies.
The BCS terms are responsible for most of the
real mass of the mesons.
The naive coupled channel terms yield the imaginary
mass (width) of the mesons. They also produce a
real contribution to the mass of mesons, however
this is canceled by the remaining coupled channel
terms, of the second and third lines in eq.
(\ref{CS3 kernel}), which readjust the spectrum 
in order to recover a vanishing mass for
the pion in the chiral limit. 
\section{Conclusions}
In this model we find that
$f_0(980)$ and $a_0(980)$ are the light $q \bar q$ scalars.
We also find that the quark $\langle \psi \bar \psi \rangle$
condensate is reduced
$5\% \rightarrow 50\%$ due to coupled channels.
\par
We also prove general model independent results.
The systematic real mass shift of the spectrum which might
be due to coupled channels cancels out. We also show
that there is a negative feedback:
the coupling $f_0 \rightarrow \pi \pi$
suppresses chiral symmetry breaking
and chiral symmetry breaking suppresses 
the coupling $f_0 \rightarrow \pi \pi$.
\par
Presently the link to QCD is being studied
with N. Brambilla, J.E. Ribeiro and A. Vairo,
and similar techniques are now being applied to
the nucleon with  S. Cotanch, Felipe, R. Fernandes,
J.E. Ribeiro and E. Swanson.
\section*{Acknowledgments}
I am very grateful to J. E. Ribeiro for long discussions and
suggestions since 1989 on the pion mass problem in relation with
coupled channels, the BCS mechanism and Ward identities.
I also thank B. Clerbaux, B. Cox, J. Phillips and M. Costa
for skiing with me at Moriond.

\end{document}